\documentclass[letter]{aa}  

\usepackage{graphicx}
\usepackage{txfonts}
\usepackage{enumitem}
\usepackage[breaklinks, colorlinks, citecolor=blue]{hyperref}
\usepackage{multirow}
\usepackage{natbib}
\usepackage{gensymb} 
\usepackage{stfloats}
\newcommand\arcdeg{\mbox{$^\circ$} }

\begin{document} 

   \title{Faraday depth similarities across scales with LoTSS \& DRAGONS}

   \titlerunning{LoTSS and DRAGONS}
   \authorrunning{Ordog et al.}
  
   \author{
            Anna Ordog\inst{\ref{uwo},\ref{ubco},\ref{drao}}\thanks{aordog@uwo.ca}
            \and
            Rebecca A. Booth\inst{\ref{ucalgary}}
            \and
            Ana Erceg\inst{\ref{ruderboskovic}}
            \and
            Vibor Jeli\'{c}\inst{\ref{ruderboskovic}}\thanks{vibor@irb.hr} 
            \and\\
            Jo-Anne C. Brown\inst{\ref{ucalgary}}
            \and
            Marijke Haverkorn\inst{\ref{radboud}}
            \and
            Alex S. Hill\inst{\ref{ubco},\ref{drao}}
            \and
            T.L. Landecker\inst{\ref{drao}}
          }

   \institute{
            Department of Physics \& Astronomy, University of Western Ontario, 1151 Richmond Street, London, ON, N6A 3K7, Canada
            \label{uwo}
        \and
            Department of Mathematics, Physics, \& Statistics, University of British Columbia Okanagan, Kelowna, BC, Canada\label{ubco}
        \and
            Dominion Radio Astrophysical Observatory, Herzberg Research Centre for Astronomy and Astrophysics, National Research Council Canada, 717 White Lake Road, Kaleden, BC V0H 1K0, Canada\label{drao}
        \and
            Department of Physics and Astronomy, University of Calgary, 2500 University Drive NW, Calgary, Alberta, T2N 1N4, Canada\label{ucalgary}
        \and
            Ruđer Bošković Institute, Bijenička cesta 54, 10 000 Zagreb, Croatia
            \label{ruderboskovic}
        \and
            Department of Astrophysics/IMAPP, Radboud University, P.O. Box 9010, 6500 GL Nijmegen, The Netherlands\label{radboud}
        }

  \abstract
   {Faraday rotation of diffuse Galactic synchrotron emission is a powerful tracer of the complex, magnetised interstellar medium (ISM), whose structures span a wide range of spatial scales, requiring both interferometric and single-antenna broadband radio polarisation observations for full characterisation. We compare Faraday rotation in the interferometric LOw-Frequency ARray Two-Metre Sky Survey (LoTSS; 120–168 MHz) and the single-antenna Dominion Radio Astrophysical Observatory Global Magneto-Ionic Medium Survey of the Northern Sky (DRAGONS; 350–1030 MHz), which are complementary in their sensitivity to spatial and Faraday-depth scales. We calculate first moments (M1) of polarised intensity versus Faraday depth, producing polarised-intensity–weighted mean Faraday depth maps of the regions common to both surveys. These maps show remarkable agreement between the surveys despite the lack of overlap in frequency or spatial-scale coverage. In the northern Galactic region, the M1 maps are morphologically similar with only small spatial shifts in the boundaries between positive and negative M1, and strong pixel-by-pixel correlation. In the southern Galactic region, both surveys trace the Faraday-depth gradient with Galactic longitude previously identified in LoTSS. Faraday depth spectra show consistent numbers and locations of peaks for more than half of the pixels. The strong structural similarity between the surveys, demonstrated by computing structure functions, suggests coupling across spatial scales in the magnetised ISM, enabling both interferometric and single-antenna observations to trace the same features. Instances of differences point to ISM configurations where observational effects such as depolarisation dominate or where this coupling breaks down due to local physical conditions.}

   \keywords{ Magnetic fields, polarisation, ISM: magnetic fields, ISM: structure, local interstellar matter, Radio continuum: ISM}

   \maketitle

\section{Introduction}
\label{Intro}
Understanding the spatial scales and line of sight (LOS) distances probed by telescopes with different characteristics is a crucial aspect of observational studies of the magnetised, ionised interstellar medium (ISM) by means of diffuse Galactic synchrotron emission Faraday rotation. Relevant considerations include frequency coverage, angular resolution, and type of instrument (single-antenna telescope or interferometer array). For diffuse polarisation and Faraday rotation, the effects of these parameters are not as well understood as they are for total intensity imaging \citep[e.g.,][]{Ordog25}. Insights from datasets differing in their observing parameters will help inform choices for future surveys and the development of analysis methods to interpret polarisation data in cross-scale Galactic magnetism studies.

Linearly polarised emission undergoes Faraday rotation of its polarisation angle by an amount proportional to the LOS integral of the LOS component of the magnetic field scaled by the electron density along the propagation path, and is quantified by the `Faraday depth' (FD),\footnote{While Faraday depth depends on the distance propagated by polarised emission, `depth' here does not map directly to a spatial depth.} also denoted by $\phi$. With the diffuse ISM synchrotron emission as the source of polarised signal, each LOS through the Galactic volume is characterised by a spectrum of FDs, the Fourier transform of the complex polarised emission versus wavelength ($\lambda$) squared \citep{brentjens05}. The recovered FD spectrum is the true spectrum convolved with the Rotation Measure Spread Function (RMSF), and the FD resolution, $\delta\phi$, is limited by the width of the RMSF's main lobe, which is inversely proportional to the $\lambda^2$ coverage \citep{brentjens05}. Consequently, lower frequency observations generally provide better $\delta\phi$. On the other hand, some scenarios of mixed synchrotron emission and Faraday rotation yield extended structures in the FD domain, which can only be recovered with the inclusion of observations at high frequencies \citep[e.g.][]{jelic15}. The widest detectable structure in an FD spectrum is inversely proportional to the smallest $\lambda^2$ observed.

\begin{figure*}[ht]
   \centering
   \includegraphics[width=\hsize]{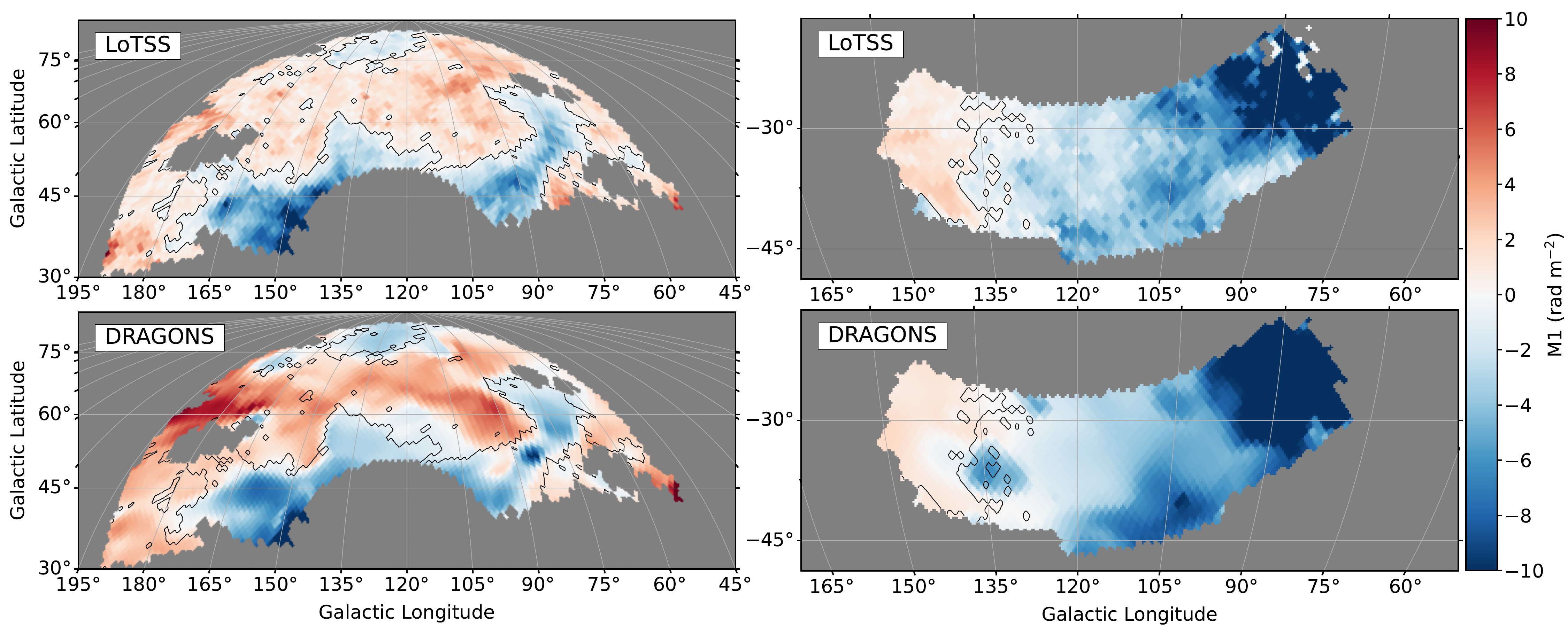}
      \caption{First moment (M1) maps for LoTSS (\textit{top}) and DRAGONS (\textit{bottom}) for the region covered by the LoTSS northern (\textit{left}) and southern (\textit{right}) mosaics. The contours on all panels trace the M1=0~rad~m$^{-2}$ boundary in LoTSS. Apart from the outline of the mosaic, the masked out (grey) regions correspond to low signal-to-noise regions in either LoTSS or DRAGONS.}\label{fig:M1_all}
\end{figure*}

In addition to FD resolution and sensitivity, a crucial consideration is the related effect of depolarisation, whereby complex polarisation is combined destructively at the telescope receiver, reducing the amount of observed emission. For mixed emission and Faraday rotation, emission from different distances rotates by different amounts, producing depth depolarisation. Small-scale variations across the telescope beam in FD or intrinsic (emitted) polarisation angle produce beam depolarisation, which can limit the distances a given instrument probes, described as the polarisation horizon \citep{Uyaniker03,Hill18}. Since the beam encompasses increasing physical volume with LOS distance, spatial fluctuations in polarisation produce more beam depolarisation at larger distances, assuming similar amplitude and scales of turbulence in the magnetised medium along the LOS. Therefore, a telescope with higher angular resolution generally probes farther distances before the signal is depolarised, which can lead to higher absolute values of detected FD in the presence of a coherent, large-scale field along the LOS. 

We compare two datasets: (i) the DRAO GMIMS Of the Northern Sky (DRAGONS) single-antenna survey and (ii) the interferometric LOw Frequency ARray (LOFAR) Two-metre Sky Survey (LoTSS) Faraday rotation data. These surveys probe markedly different parameter spaces in frequency and angular scale coverage, yet reveal remarkable similarities in their diffuse Faraday rotation morphology.

\section{Data preparation}\label{sec:data}
DRAGONS \citep{Ordog26} is a single-antenna northern sky polarisation dataset produced using the 15~m telescope at the Dominion Radio Astrophysical Observatory (DRAO) as the `Low Band North' 350--1030 MHz component of the Global Magneto-Ionic Medium Survey \citep[GMIMS;][]{Wolleben2019,wolleben21,Sun2025}. The DRAGONS FD cube has an RMSF width of 6~rad~m$^{-2}$ with a nominal sensitivity to structures as wide as 40~rad~m$^{-2}$ \citep[following][]{brentjens05}. 

LoTSS is a 120--168 MHz interferometric survey of the northern sky \citep{shimwell17} with the LOFAR High Band Antennas \citep{vanhaarlem13}. We use two FD cubes of mosaicked Galactic regions: a northern \citep{erceg22} and a southern region \citep{erceg24}, with both an RMSF width and a maximum FD scale of $\sim1~{\rm rad~m^{-2}}$. Polarised point sources were not subtracted from the LoTSS data since their low spatial density \citep{OSullivan2023} makes them unlikely to significantly affect the analysis of diffuse Galactic emission. 

We provide technical details of both surveys in Appendix~\ref{app:data}. Starting with the published FD cubes, we binned and masked the datasets to facilitate a direct pixel-by-pixel comparison. We down-sampled both the LoTSS and DRAGONS FD Polarised Intensity (PI) cubes to a $\sim1^{\circ}$ spacing on a Healpix grid of \texttt{nside}=64. This adequately samples the $3.6^{\circ}$ resolution DRAGONS data, and is a negligible down-sampling of the published cubes, not averaging any independent pixels. For LoTSS, this is a significant degradation from its $5.5'$ resolution, but this method retains the signal needed to compare large-angular-scale structures between the two surveys. Convolving LoTSS Stokes $Q$ and $U$ to the DRAGONS angular resolution results in almost complete depolarisation of the signal because of positive and negative $Q$ and $U$ fluctuations on scales smaller than the nominal $\sim1\arcdeg$ largest scales probed by LoTSS \citep{shimwell17}.\footnote{Convolving LoTSS Stokes $Q$ and $U$ to the $40'$ resolution of GMIMS High Band North in \cite{erceg22} did not exceed this limit.} We did not attempt to reconcile the differing FD resolution of the surveys by smoothing LoTSS in FD space, preferring to compare the native spectra. We masked the cubes to the two LoTSS mosaics, further masking low signal-to-noise areas in either survey.

\section{First moment comparison}\label{sec:results}
For a comparison of the spatial structures in Faraday rotation revealed by LoTSS and DRAGONS, we use first moment (M1) maps of the FD cubes, the polarised-intensity-weighted averages of the FDs in each spectrum. M1 has been used extensively to analyse large-scale, average LOS magnetic field patterns traced by Faraday rotation of diffuse synchrotron emission \citep[e.g.][]{dickey19,Dickey2022,erceg22,erceg24,Raycheva26,Booth26}. For a two-dimensional comparison, M1 maps are a good alternative to peak FD maps, because the weighted average retains information about the full spectrum that is lost when simply finding the peak. In the case of a complex LOS, both surveys may detect most of the multiple FD components, but the relative PI of the peaks may differ between the surveys, leading to apparent disagreement. We use M1 (calculated over $|\phi|<50$~rad~m$^{-2}$; Fig.~\ref{fig:M1_all}) for our main analysis since it is less subject to this effect, and show the poorer agreement between DRAGONS and LoTSS peak FD in Fig.~\ref{img:FDmatchpeakmap}. 

\begin{figure}[ht]
   \centering
   \includegraphics[width=0.98\hsize]{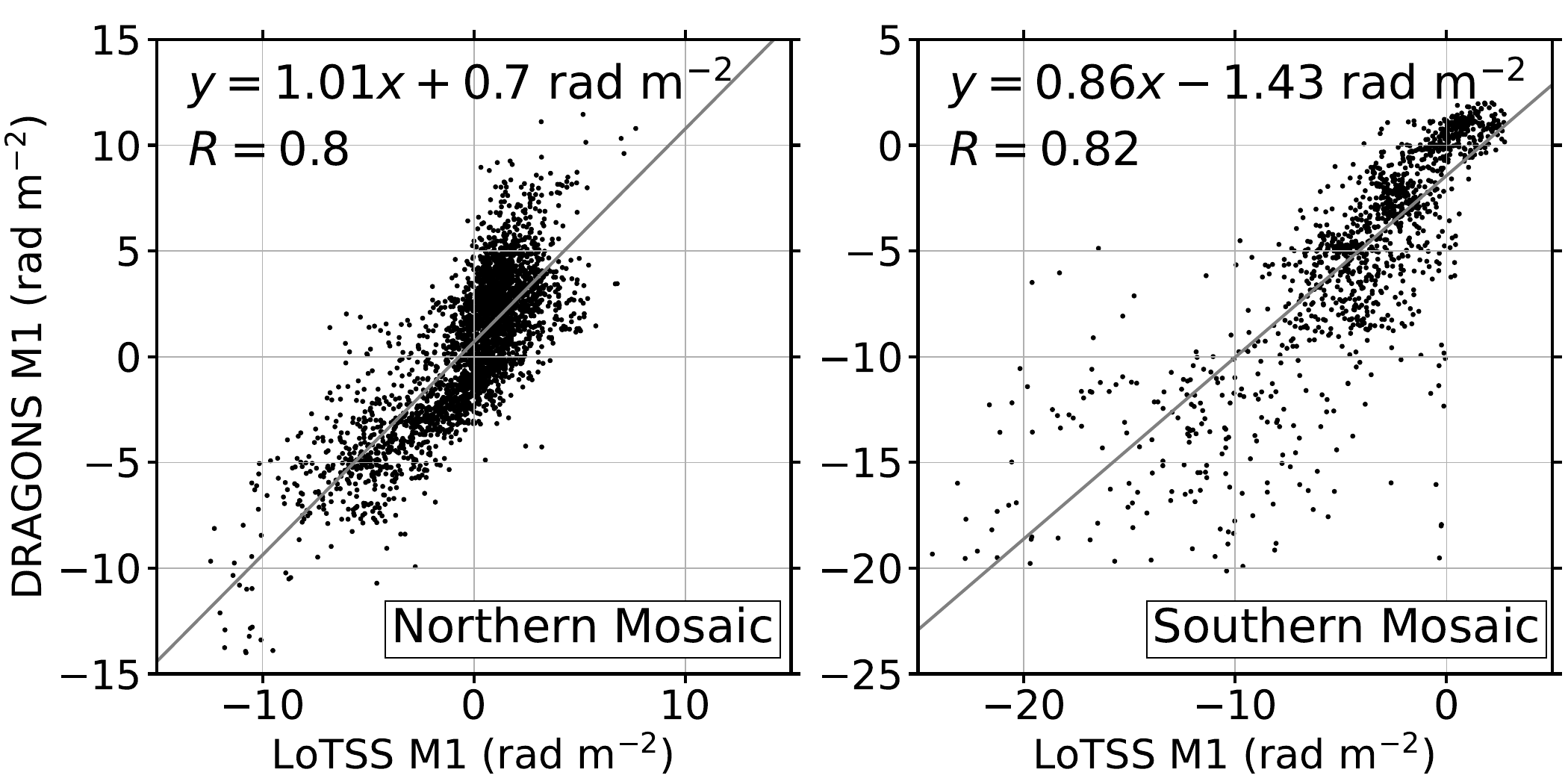}
      \caption{M1 - M1 plots, directly comparing the M1 values of DRAGONS versus LoTSS for the pixels in Fig.~\ref{fig:M1_all}. We note the strong correlation in both regions, and the good agreement in M1 values as indicated by the slopes of the best-fit lines, particularly for the northern mosaic. }  \label{img:MnM}
\end{figure}
\begin{figure}[ht]
   \centering
   \includegraphics[width=\hsize]{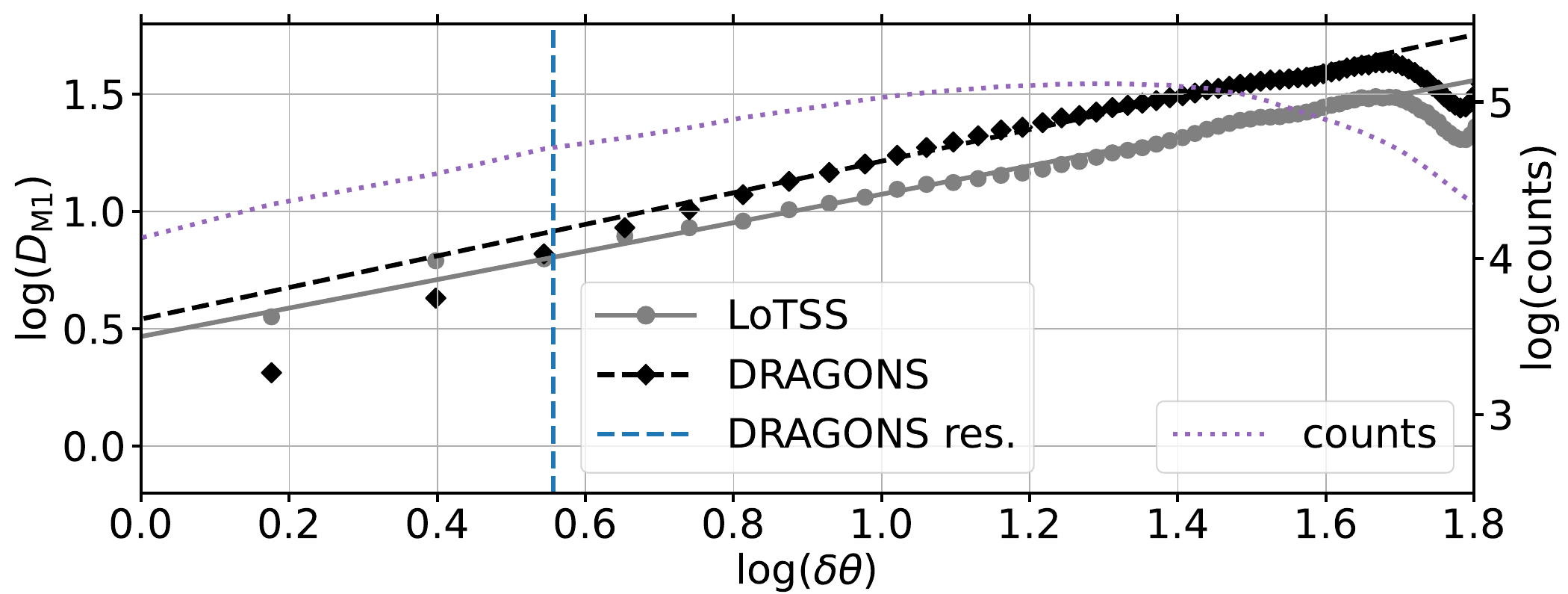}
      \caption{Second order structure functions of M1 in the combined mosaic maps for LoTSS (grey circles) and DRAGONS (black diamonds). The solid grey (black dashed) line shows the LoTSS (DRAGONS) slope fitted over $0.6<\log(\delta\theta)<1.6$. The vertical dashed blue line indicates the DRAGONS resolution and the dotted purple line indicates the counts of pixel pairs in the calculation.}  \label{img:SFmain}
\end{figure}
\begin{figure}[ht]
   \centering
   \includegraphics[width=\hsize]{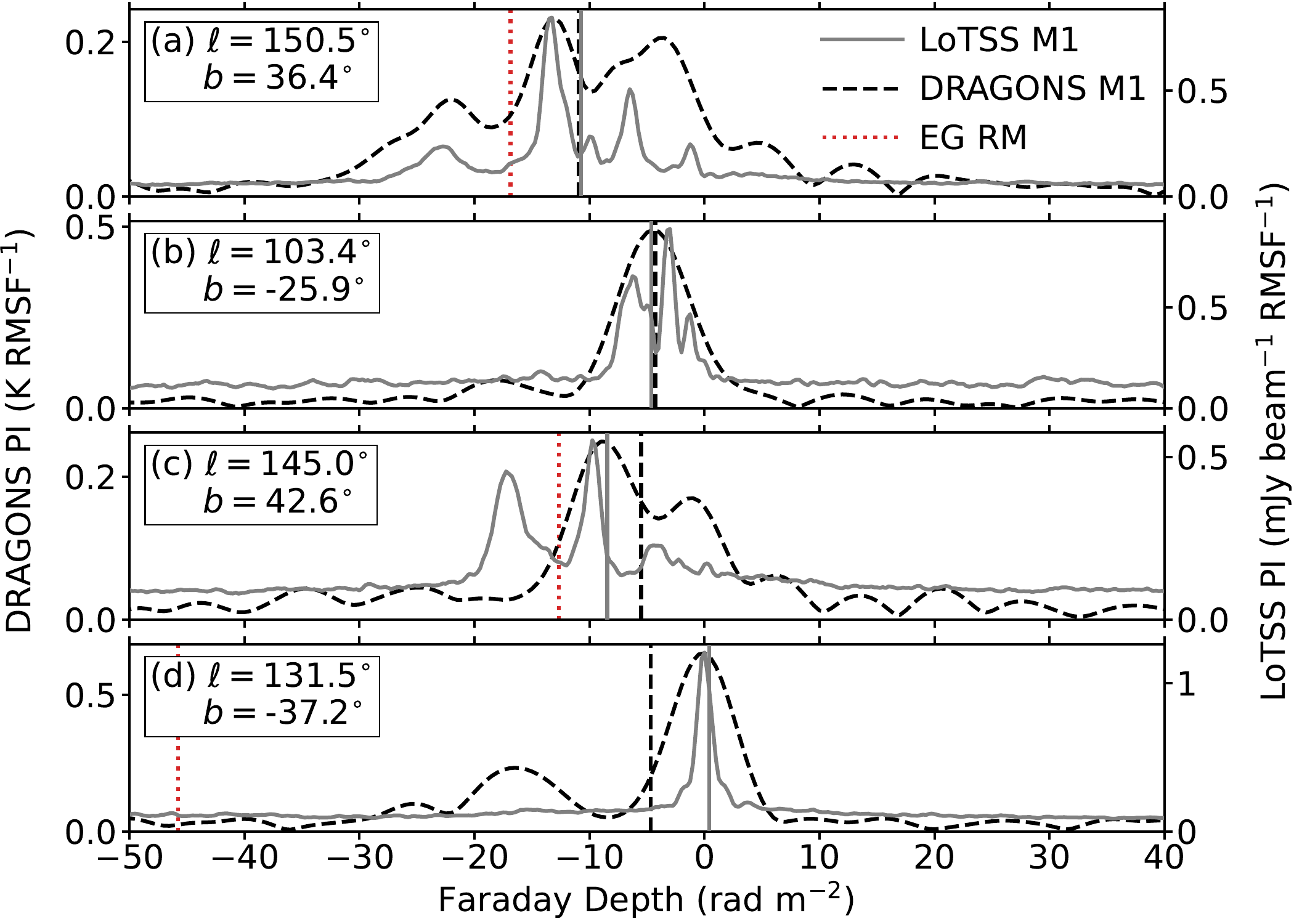}
      \caption{Examples of LoTSS (solid grey) and DRAGONS (dashed black) spectra highlighting similar and differing sight-lines with M1 values indicated by corresponding vertical lines. Red dotted lines indicate the EG RM value from the \cite{hutschenreuter22} Galactic Faraday rotation map, which in the case of panel (b) is outside the displayed $\phi$ range. Note the differing units for PI in LoTSS and DRAGONS.}\label{img:example_spectra_v2}
\end{figure}

In the northern mosaic (Fig.~\ref{fig:M1_all}, left panels) we note a strong similarity between many of the large-scale structures traced by the two data sets. The boundary between positive and negative M1 (tracing a reversal in the average LOS magnetic field component) approximately follows the upper part of Loop III, an arching synchrotron emission structure along Galactic latitude $b \approx 60^\circ$ \citep[noted for LoTSS in][]{erceg22}. In DRAGONS there is only a slight northward shift of this boundary compared to LoTSS near $\ell=120\arcdeg$, and a few patches of positive M1 where LoTSS has negative M1 near ($\ell,b$)~=~($170\arcdeg,50\arcdeg$) and ($95\arcdeg,50\arcdeg$). If we consider FD peaks instead of M1, taking the LoTSS FD feature for each LOS that most closely matches the DRAGONS peak FD (Fig.~\ref{img:FDmatchpeakmap}, bottom panels) the positive-negative boundary agrees with DRAGONS. Thus the shift seen in M1 may be attributed to the broader DRAGONS RMSF blending the spectral features. In the southern mosaic (Fig.~\ref{fig:M1_all}, right panels), the large-scale M1 gradient noted in LoTSS by \cite{erceg24} is also evident in DRAGONS, although in DRAGONS there is a patch of more strongly negative M1 near $(\ell,b)=(135\arcdeg, -40\arcdeg)$.

For a pixel-by-pixel comparison, we plot DRAGONS versus LoTSS M1 in Fig.~\ref{img:MnM}. We note the remarkably strong Pearson correlation values for both mosaics, and the excellent agreement in M1 values, particularly in the northern mosaic, highlighted by the near-unity slope. We discuss the small population of pixels in the northern mosaic for which DRAGONS has larger positive M1 values in Appendix~\ref{app:pol-bias}. In the southern mosaic, there is a slight overall negative offset in the DRAGONS M1 and increased scatter for high negative M1 values. We visualize this further by binning both datasets by longitude in Fig.~\ref{img:gradient}. This emphasizes the strong presence of the southern M1 gradient in both datasets but also reveals an increasing deviation toward stronger M1 values in DRAGONS moving from high to low Galactic longitudes. There are shallower M1 gradients present in the northern mosaic, attributable to Galactic-scale magnetic structures \citep[e.g.,][]{Dickey2022,Booth26}.

Due to the interferometric nature of LoTSS, some large-angular-scale diffuse emission structures are filtered out in individual frequency channel PI maps. However, variations in FD distribution across the sky can manifest structures on angular scales larger than the fundamental interferometric limits, as evidenced by similar patterns appearing in LoTSS and the large-scale-dominated DRAGONS M1 maps. We quantify these similarities across spatial scales using second order M1 structure functions, calculated as $D_{\text{M1}}(\delta\theta)= \langle [\text{M1}(\theta)-\text{M1}(\theta+\delta\theta)]^2 \rangle_{\theta}$, where $\theta$ is the angular position of each pixel in the maps with the average taken over all $\theta$ in the northern and southern mosaics combined (Fig.~\ref{img:SFmain}). The angular separations are $\delta\theta=\cos^{-1}(\hat{r}_1\cdot\hat{r}_2)$, where $\hat{r}_i$ is the normalized vector corresponding to $\theta_i$. As expected, the DRAGONS $D_{\text{M1}}$ slope steepens toward the $1\arcdeg$ map sampling below the DRAGONS $3.6\arcdeg$ angular resolution (vertical dashed line), while the LoTSS slope remains roughly constant across this scale. We calculated the slopes over the range of $\delta\theta$ where $D_{\text{M1}}$ is approximately linear as $0.61\pm0.01$ for LoTSS and $0.67\pm0.01$ for DRAGONS. Although the values do not strictly agree to within the uncertainties, we consider this to indicate moderately strong agreement from $\sim4\arcdeg$ up to $\sim40\arcdeg$ scales, emphasizing that both surveys trace a wide range of large-scale structure in FD M1. This highlights the strength of Faraday rotation analyses of interferometric data in recovering coherent, large-scale FD patterns from detected small-scale spatial fluctuations in diffuse, complex polarised emission \citep[e.g.,][]{Haverkorn04, Ordog17, erceg22, erceg24}.

\section{Faraday Depth spectra}\label{sec:discussion}
We have used M1 as a proxy to compare the Faraday rotation probed by each of the two surveys. We now examine FD spectra for sample sight-lines (Fig.~\ref{img:example_spectra_v2}), which hint at possible ISM structures that, combined with characteristics of the surveys, may cause the similarities and differences. For these sight-lines we also show extragalactic (EG) Faraday Rotation Measure (RM) values from the \cite{hutschenreuter22} map. For some sight-lines the FD values in the diffuse emission spectrum are comparable to the corresponding EG RM, while in other cases the EG sources and diffuse emission appear to probe different volumes.\footnote{A caveat: EG RMs may have large errors when derived from few frequency channels, and may be interpolated from sparse spatial sampling.} In many cases EG RMs alone appear to be a poor indicator of the FD distribution along the LOS, and valuable information is added by considering the diffuse emission to investigate, for example, observed complexity in EG FD spectra attributable to mixed-in diffuse Galactic emission \citep{Ranchod2024}. Future work will include developing techniques to classify the types of spectra in the $\sim4000$ lines of sight across the two mosaics, which will in turn allow for the identification of different magneto-ionic distributions throughout the Galaxy.

For approximately one third (two thirds) of the maps by area, the LoTSS and DRAGONS M1 values agree to within 1~rad~m$^{-2}$ (2~rad~m$^{-2}$). Given the expected differences in depolarisation between the surveys, this agreement implies a correlation of the magnetised structures along the LOS. For many lines of sight the spectra in both surveys have multiple peaks, with 2 to 4 dominant peaks agreeing in FD. The LOS in Fig.~\ref{img:example_spectra_v2}(a), from the strongly negative M1 region near the bottom of Loop III, features several negative FD peaks whose locations agree between the two surveys (see Fig.~\ref{img:FDmatchpeakmap}) though their relative amplitudes differ.
 
Another scenario of M1 agreement is shown in Fig.~\ref{img:example_spectra_v2}(b), where the broader DRAGONS RMSF spans multiple closely spaced spectral components detected by the higher FD resolution of LoTSS. Small LOS fluctuations in electron density or magnetic field cannot be observed with the poorer FD resolution of DRAGONS. Single (Faraday thin) peaks in DRAGONS may also arise from extended FD structures, produced by well-mixed synchrotron emission and Faraday rotation along the LOS. LoTSS, with poor sensitivity to extended FD structures, will detect only their edges due to depth depolarisation.

Some sight-lines reveal an extra FD peak seen only in LoTSS. In Fig.~\ref{img:example_spectra_v2}(c) both surveys have multiple FD features over $-10<\phi<0$~rad~m$^{-2}$, but LoTSS has an additional peak near $-18$~rad~m$^{-2}$. The EG sources contributing to the \cite{hutschenreuter22} map at this position reveal sufficient spread in Faraday rotation to produce beam depolarisation of distant diffuse emission in the $3.6\arcdeg$ DRAGONS beam through the resulting spread in polarisation angle, making this FD peak undetectable. The smaller LoTSS beam, subject to less beam depolarisation, may probe to a more distant polarisation horizon. Averaging the LoTSS PI spectra in $\sim1\arcdeg$ pixels allows detection of this peak despite the spatial fluctuations. In a small fraction of sight-lines point-source contamination may be present in LoTSS, appearing as low-PI peaks matching the FD of the corresponding EG source. This is not a concern for the example in Fig.~\ref{img:example_spectra_v2}(c), as there are no LoTSS point sources within the pixel, and the $-18$~rad~m$^{-2}$ feature is spatially coherent over adjacent pixels. The smaller $|\phi|$ features in both surveys may be associated with more nearby Faraday screens, illuminated by the local diffuse emission.

We also observe sight-lines where LoTSS lacks an FD peak present in DRAGONS. In the southern mosaic, in a region several degrees wide near $(\ell,b)=(135\arcdeg,-40\arcdeg)$, only DRAGONS has a secondary negative FD peak in addition to the $\sim0$~rad~m$^{-2}$ peak appearing in both surveys (Fig.~\ref{img:example_spectra_v2}(d)). This may indicate large-scale magnetic structure to which DRAGONS is sensitive and lacks the coupling to small scales that would make it visible in LoTSS. While in most cases the large-scale physics appears to be imprinted on the small angular scales detectable by LoTSS, there may be FD features arising from magnetic field and electron density configurations that are uniform on scales larger than the $\sim1\arcdeg$ angular-scale LoTSS sensitivity.

\section{Conclusions}\label{sec:conclusions}
We have demonstrated that similar large-angular-scale Faraday rotation structures are traced in diffuse Galactic synchrotron emission mapped by the DRAGONS and LoTSS polarisation surveys. This is significant given the limitations of the LoTSS interferometric data in sensitivity to large angular scales in PI, and the expected depolarisation arising from the large beam of the single-antenna telescope used for DRAGONS. The FD M1 maps reveal similarities in spatial structures, while the spectra confirm that for most sight-lines the surveys probe a similar ISM volume or indicate a correlation of the magnetised structures along the LOS. Appropriate metrics for characterising and classifying the significant Faraday complexity in both surveys need to be developed, and will be the focus of future work including simulations of the underlying ISM configurations \citep[e.g.][]{Basu2019}.

This study highlights the ability of interferometric radio polarisation data to trace large-angular-scale Faraday rotation patterns, pointing to a connection between small- and large-scale magnetic field structures in the ISM. Comparing interferometric and single-antenna diffuse emission FDs will also be possible with the Polarisation Sky Survey of the Universe's Magnetism \citep[POSSUM;][]{Gaensler2025} and the in-progress POSSUM EMU (Evolutionary Map of the Universe) GMIMS All-Stokes UWL (Ultra Wideband Low) Survey (PEGASUS), with overlapping frequency coverage eliminating one of the variables present in our LoTSS-DRAGONS comparison. A complete understanding of magnetic field structures across spatial scales is essential to ISM physics, making it crucial to take into account observational effects such as those we have begun to explore here.

\begin{acknowledgements}
DRAO is located on the traditional, ancestral, and unceded territory of the syilx Okanagan people. R.A.B. was supported by the Natural Sciences and Engineering Research Council of Canada (NSERC) Vanier scholarship and the University of Calgary Izaak Walton Killam Doctoral Scholarship. This work was supported by NSERC Discovery Grants (PIs: J.C.B., A.S.H., T.L.L.). We acknowledge European Union - NextGenerationEU and the COsmic Magnetism with RADio Astronomy 2024 (COMRAD2024) conference. We thank the anonymous referee for their constructive feedback on this manuscript.
\end{acknowledgements}

\bibliographystyle{aa}
\bibliography{reference_list.bib} 

\begin{appendix}
\raggedbottom
\section{Data specifics and availability}\label{app:data}

 \begin{table*}[b]
    \centering
    \caption{\centering \label{tab:surveys} Summary of the LoTSS vlow and DRAGONS survey parameters.}
    \begin{tabular}{l c c}
    \hline\hline
     & LoTSS vlow & DRAGONS \\
    \hline
    baselines & 100~m - 1.6~km & 0 - 15~m\\
    resolution & $4'-5.5'$ & 3.6\arcdeg \\
    maximum angular scales & $~\sim1\arcdeg$ & -- \\
    sky coverage & north (centre RA = $12^{\rm h}$): $\sim3100$ deg.$^2$& $-20\arcdeg<\text{dec.}<90\arcdeg$\\
                 & south (centre RA = $0^{\rm h}$): $\sim1400$ deg.$^2$& \\
    frequencies & 120 - 168 MHz & 350 - 1030 MHz \\
    channel widths & 97.6 kHz &  1 MHz\\ 
    $\lambda^2$ coverage & 3.2 - 6.25 m$^{2}$ & 0.085 - 0.73 m$^{2}$\\
    $\delta \phi$ & $\sim 1$ rad~m$^{-2}$ & $\sim 6$ rad~m$^{-2}$\\
    $\phi_{\text{max-scale}}$ & $\sim 1$ rad~m$^{-2}$ & $\sim 40$ rad~m$^{-2}$\\
    $|\phi_\mathrm{max}|$ & $\sim 190$ rad~m$^{-2}$ & $\sim 450$ rad~m$^{-2}$\\
    $\phi$ cube noise & 71 $\mu$Jy PSF$^{-1}$ RMSF$^{-1}$ & 11~mK~RMSF$^{-1}$\\
    \hline
    \end{tabular}
\end{table*}
This appendix provides a description of the two datasets used in this analysis, with further details available in \cite{Ordog26} for DRAGONS and in \cite{erceg22, erceg24, shimwell22} for LoTSS vlow. Table~\ref{tab:surveys} provides the observing parameters and characteristics of the two surveys. 

The DRAGONS single-antenna 350-1030~MHz northern sky polarisation dataset was produced using the 15~m telescope at the DRAO. It covers a declination range of $-20\arcdeg<\text{dec.}<90\arcdeg$. The angular resolution is 1.3\arcdeg at 1030~MHz, but prior to RM synthesis DRAGONS Stokes $Q$ and $U$ cubes were convolved to the 3.6\arcdeg angular resolution of the 350~MHz channel. RM synthesis and CLEAN were performed using \texttt{RM-Tools} \citep{purcell20, VanEck2026} with a CLEAN threshold of $5\sigma$. The frequency coverage yields an RMSF width of 6~rad~m$^{-2}$ with a nominal sensitivity to structures as wide as 40~rad~m$^{-2}$ \citep{brentjens05}. DRAGONS data are available on the Canadian Advanced Network for Astronomical Research (CANFAR) at
 \href{https://www.canfar.net/citation/landing?doi=25.0104}{10.11570/25.0104}. 
 
LoTSS is an interferometric survey of the northern sky \citep{shimwell17} with the LOFAR High Band Antennas \citep[HBA;][]{vanhaarlem13} covering 120--168 MHz. The two FD cubes of mosaicked Galactic regions (northern and southern) have angular resolution of $5.5'$ (39 times finer than DRAGONS) with nominal sensitivity up to $\sim1\arcdeg$ angular scales. They are based on the LoTSS Data Release 2 \citep[DR2;][]{shimwell22} very low-resolution (vlow) Stokes $Q$ and $U$ undeconvolved image cubes of calibrated pointings. The frequency coverage yields both an RMSF width and a maximum FD scale of $\sim1~{\rm rad~m^{-2}}$. RM CLEAN was not applied because the RMSF sidelobes are below 20\% and the signal-to-noise ratio is generally low, so deconvolution would have only a negligible effect. Polarised point sources were not subtracted from the LoTSS data (see Sect.~\ref{sec:data}). Mosaicked FD cubes of the LoTSS-DR2 vlow data used in this work are available on FULIR Data, the research data repository of the Ru{\dj}er Bo\v{s}kovi\'c Institute, at \href{https://urn.nsk.hr/urn:nbn:hr:241:389701}{https://urn.nsk.hr/urn:nbn:hr:241:389701} and
\href{https://urn.nsk.hr/urn:nbn:hr:241:245236}{https://urn.nsk.hr/urn:nbn:hr:241:245236}.

For downsampling the Faraday depth cubes of the two surveys to a matched gridding as described in Sect.~\ref{sec:data}, we used the \texttt{healpy} function \texttt{ud\_grade}. For both surveys we calculated M1 over $|\phi|<50$~rad~m$^{-2}$, much smaller than the maximum detectable Faraday depth $|\phi_\mathrm{max}| \gtrsim 200$~rad~m$^{-2}$ for both. For DRAGONS we used a $6\sigma$ PI threshold, and for LoTSS an adaptive threshold of the mean PI plus $5\sigma$, following the methods in \cite{Ordog26} and \cite{erceg22} for DRAGONS and LoTSS respectively.

\section{Supplementary figures}\label{app:figures}
\begin{figure}[ht]
    \centering
        \includegraphics[width=\hsize]{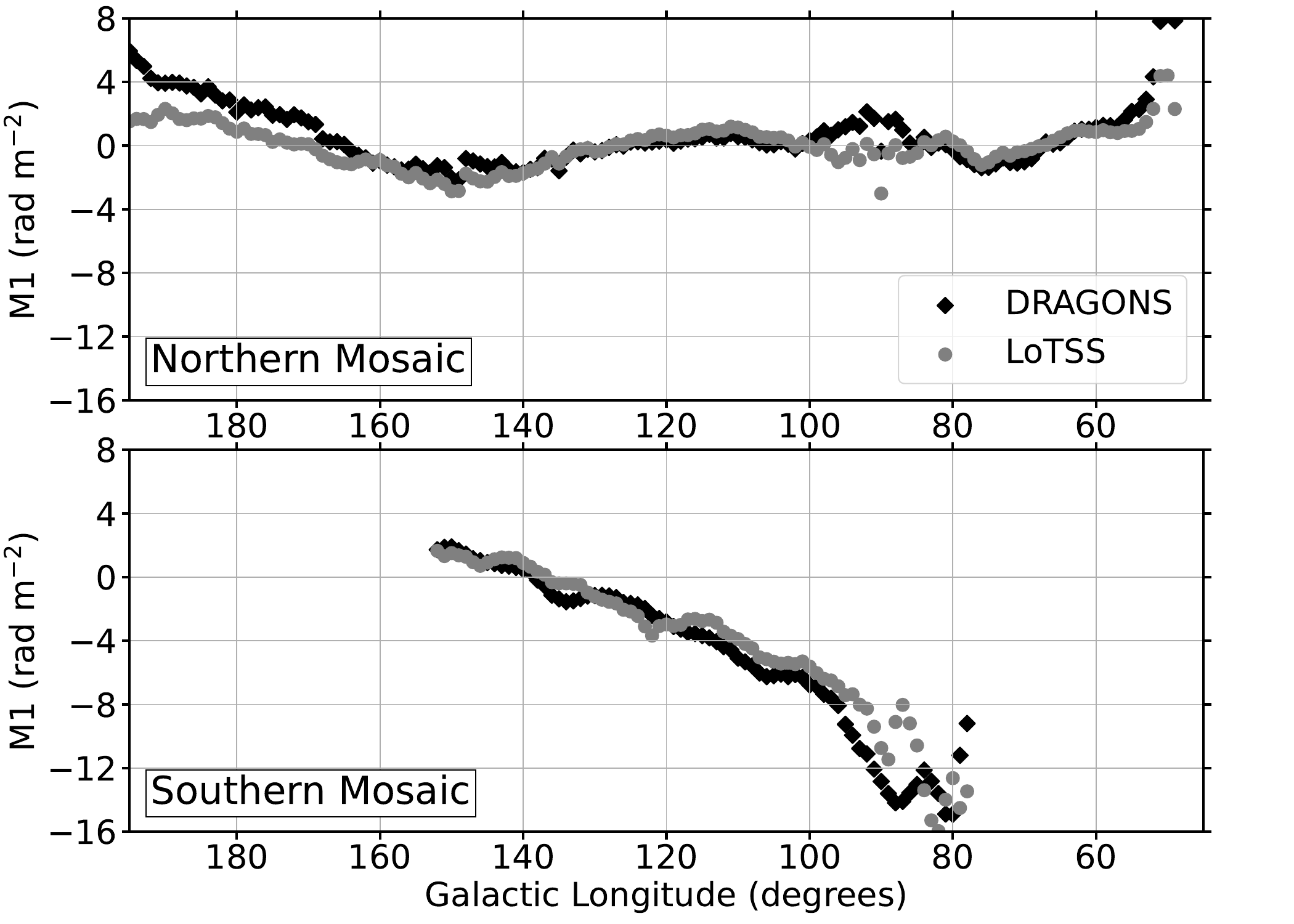}
        \caption{Northern (\textit{top}) and southern (\textit{bottom}) mosaic M1 values for LoTSS (grey circles) and DRAGONS (black diamonds) binned by Galactic longitude, highlighting the strong gradient present in both datasets in the southern mosaic.}
        \label{img:gradient}
\end{figure}

We provide additional figures highlighting the spatial patterns seen in the LoTSS and DRAGONS maps. Fig.~\ref{img:gradient} shows the M1 of both datasets binned by Galactic longitude, emphasizing the strong presence of the southern M1 gradient in both. In the northern mosaic, the gradient across $\ell>150\arcdeg$ traces the `$\sin{2\ell}$' Faraday rotation pattern identified and modelled in \cite{Dickey2022}, and the minimum at $\ell=150\arcdeg$ is likely due to the large-scale magnetic field reversal \citep[e.g.,][]{Booth26}.

\begin{figure}[t]
        \centering
        \includegraphics[width=\hsize]{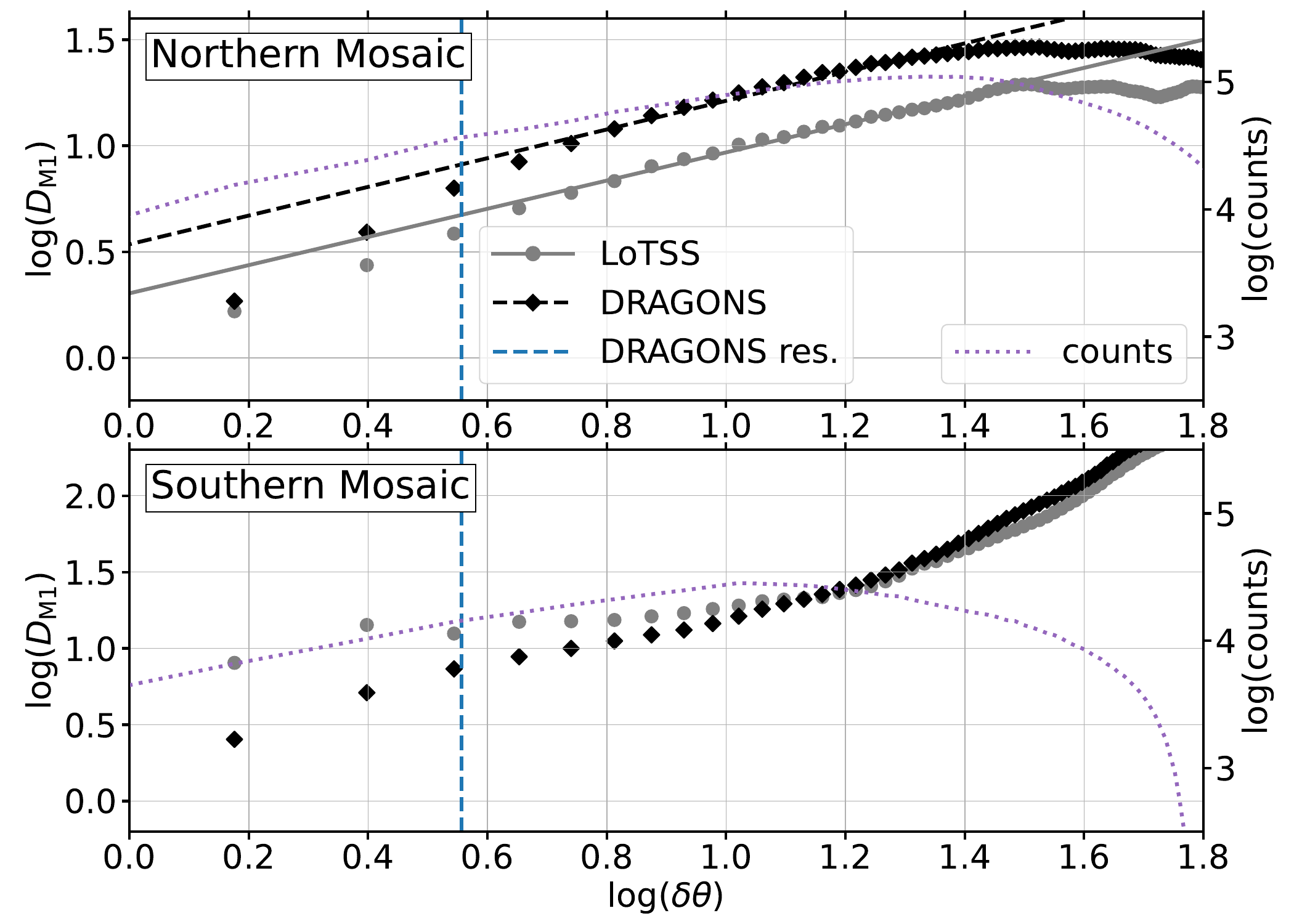}
        \caption{Second order structure functions of M1 for LoTSS (solid grey) and DRAGONS (dashed black) in the northern (\textit{top}) and southern (\textit{bottom}) mosaics. The vertical dashed blue line indicates the DRAGONS resolution and the dotted purple line indicates the counts of pixel pairs in the calculation.}
        \label{img:structurefunction}
\end{figure}

Fig.~\ref{img:structurefunction} shows the second order structure functions for DRAGONS and LoTSS M1 values in the northern and southern mosaics separately. As in Fig.~\ref{img:SFmain}, the dotted purple line indicates the number of pixel pairs that go into each $D_{\text{M1}}(\delta\theta)$ calculation. Note that these counts drop rapidly when approaching the angular scale of the mosaic masks. As for the combined mosaics, in each of the separate mosaics the DRAGONS $D_{\text{M1}}$ slope steepens toward the $1\arcdeg$ map sampling below the DRAGONS $3.6\arcdeg$ angular resolution, while the LoTSS slope $D_{\text{M1}}$ remains roughly constant across this scale. In the northern mosaic, there appears to be an offset between the LoTSS and DRAGONS $D_{\text{M1}}$, consistent with the generally slightly larger $|\text{M1}|$ values in DRAGONS. In the southern mosaics, both LoTSS and DRAGONS $D_{\text{M1}}$ feature a steepening toward the scale of the width of the southern mosaic mask ($\sim65\arcdeg$ wide), highlighting the large-scale gradient in both. 
\begin{figure*}[b]
   \centering
   \includegraphics[width=\hsize]{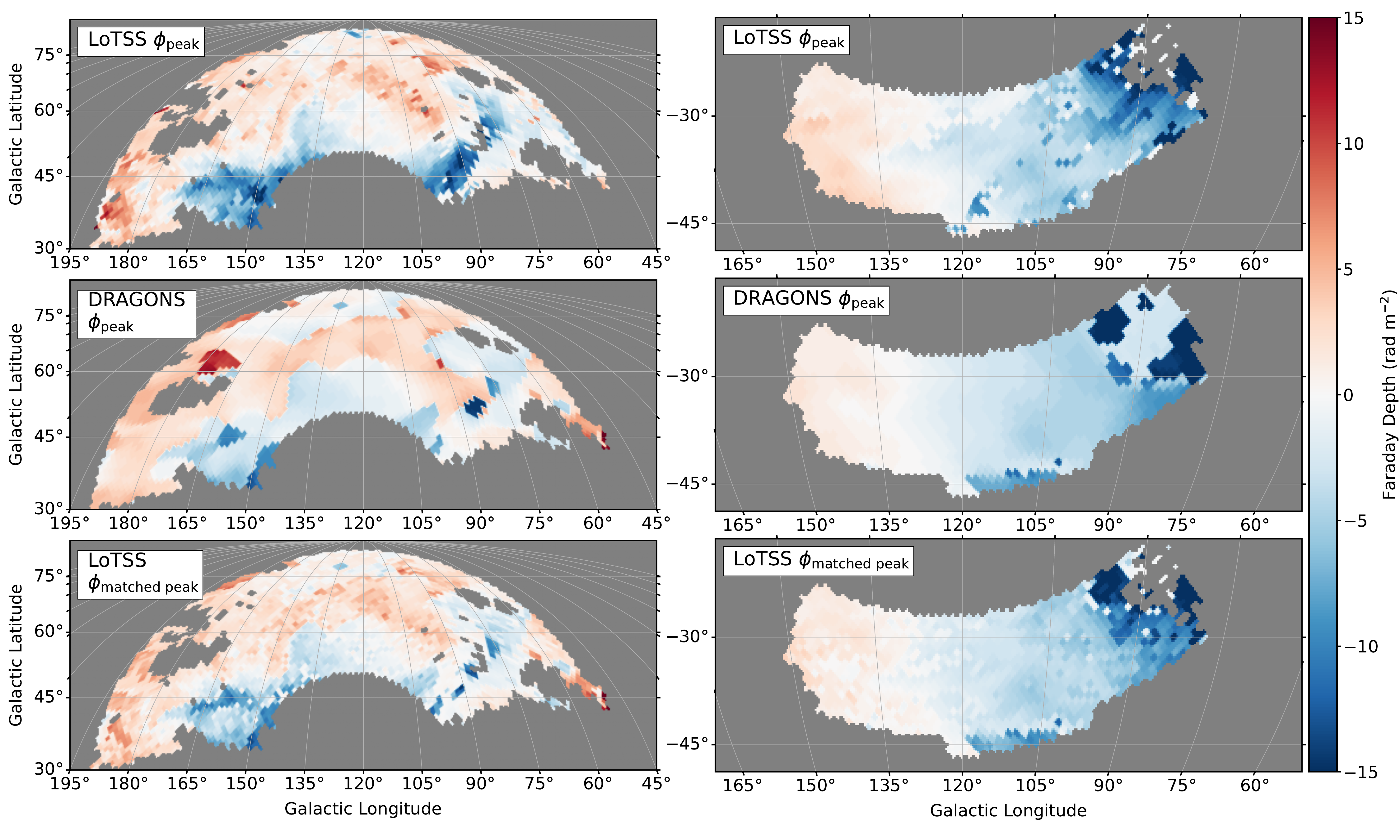}
      \caption{Peak FD maps for LoTSS and DRAGONS. \textit{Top}: FDs corresponding to the highest PI peaks in each LoTSS spectrum. \textit{Middle}: FDs corresponding to the highest PI peaks in each DRAGONS spectrum. \textit{Bottom}: The FD of the LoTSS peak closest in value to the highest peak DRAGONS FD for each LOS.}
      \label{img:FDmatchpeakmap}
\end{figure*}

The structure function slopes over the range of $\delta\theta$ where $D_{\text{M1}}(\delta\theta)$ is approximately linear ($0.6<\log(\delta\theta)<1.4$) are $0.66\pm0.004$ and $0.68\pm0.01$ for LoTSS and DRAGONS respectively in the northern mosaic, indicating similar agreement as in the combined mosaic structure function slope. We do not calculate slopes for the southern mosaic structure functions separately, as their profiles are dominated by the steepening resulting from the gradient. Future work will include a more refined analysis to determine the exact scales over which the two surveys can be said to definitively agree in the Faraday rotation structures they trace, and how this agreement may vary between different regions on the sky, along with a comparison to previous literature on RM structure functions \citep[e.g.,][]{haverkorn08}.

Fig.~\ref{img:FDmatchpeakmap} shows the FD values at the peaks in the LoTSS and DRAGONS spectra. This provides complementary information to the M1 maps shown in Fig.~\ref{fig:M1_all}. In the case of Faraday simple emission (a spectrum dominated by a single Gaussian), M1 and peak FD should be identical. The top two rows of panels show the FDs corresponding to the highest PI peaks in each pixel for LoTSS and DRAGONS ($\phi_{\text{peak}}$). We note the reduced agreement between the surveys compared to the M1 maps. To explore this further, for each pixel in the maps we found the spectral feature in LoTSS most closely matching the DRAGONS $\phi_{\text{peak}}$ at that pixel ($\phi_{\text{matched peak}}$). The result is shown in the bottom row of Fig.~\ref{img:FDmatchpeakmap}. There is increased resemblance between DRAGONS $\phi_{\text{peak}}$ and LoTSS $\phi_{\text{matched peak}}$ compared to LoTSS $\phi_{\text{peak}}$. This suggests that while the relative power at multiple FD features differs between the surveys, spectral features common to both exist over most of the mapped region, forming coherent, matching structures.

\section{Sources of discrepancy between LoTSS and DRAGONS M1}\label{app:pol-bias}
In Sect.~\ref{sec:results} we noted a small population of pixels for which DRAGONS has larger positive M1 values than LoTSS, as seen in Fig.~\ref{img:MnM}. For some of these lines of sight, the LoTSS and DRAGONS FD spectra also reveal differing numbers of peaks or disagreement in the FD values of the peaks. A subset of these may be attributed to instrumental polarisation in LoTSS, in the form of total intensity leakage of bright compact sources into Stokes $Q$ and $U$, producing spurious features at low $|\phi|$ values, which then tend to pull the M1 toward zero. This instrumental effect may be stronger in the binned LoTSS pixels compared to the native resolutions maps, since leakage from a single compact source can contaminate the larger, binned pixel in which it is located. Future work will address this by masking out bright source pixels prior to binning. Additionally, it is possible that for some lines of sight spurious features are present in the LoTSS spectra as a result of not applying RM CLEAN or other methods of deconvolving the RMSF, which can also affect M1 maps but not peak FD maps.

\begin{figure}[ht]
   \centering
   \includegraphics[width=\hsize]{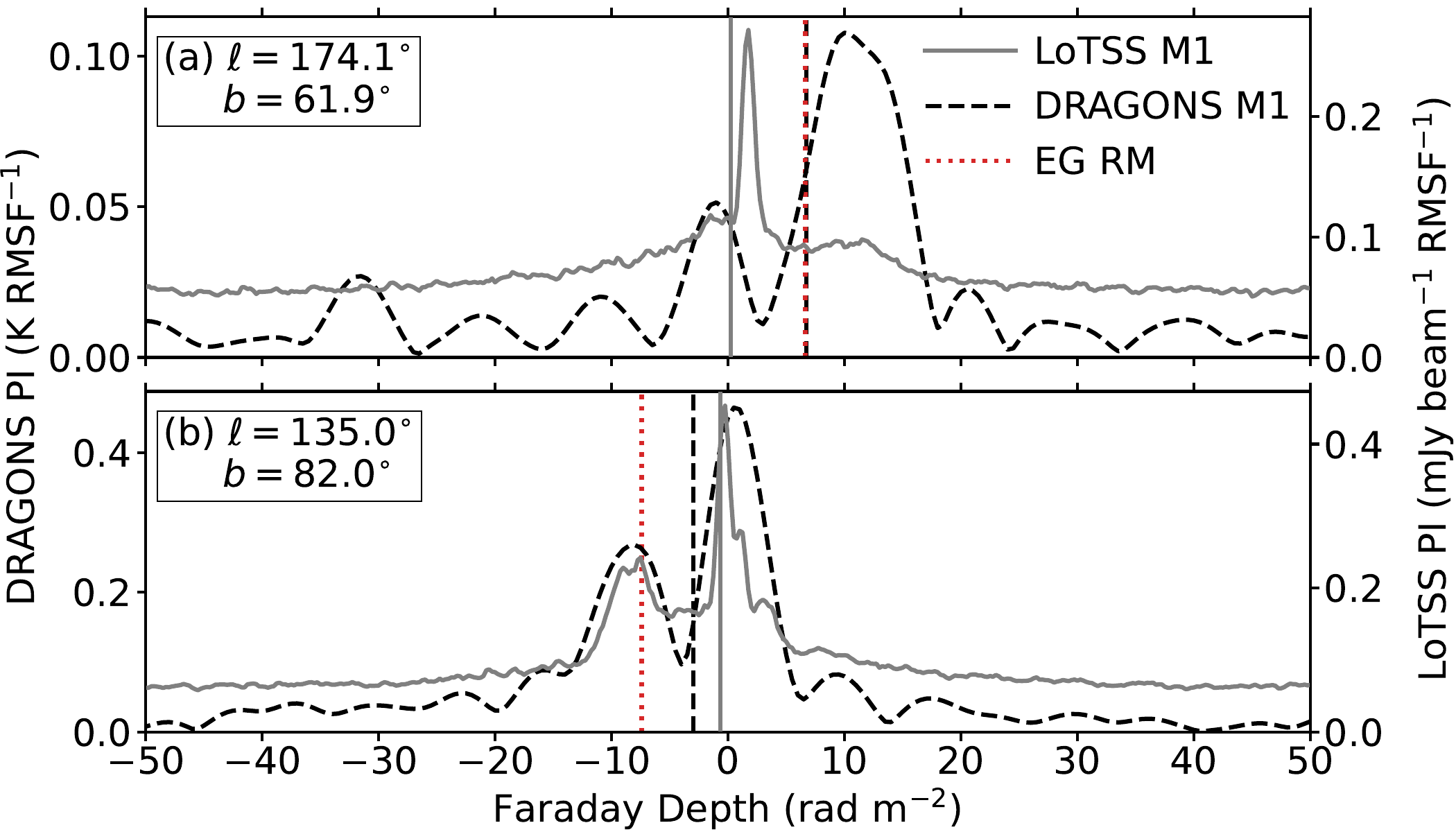}
      \caption{Examples of LoTSS (solid gray) and DRAGONS (dashed black) spectra demonstrating the effects of polarisation bias on M1 values (indicated by corresponding vertical lines). Red dotted lines indicate the EG RM value from the \cite{hutschenreuter22} Galactic Faraday rotation map. Note the differing units for PI in the two surveys, and the polarisation bias offset in LoTSS PI.}   \label{img:example_spectra_polbias}
\end{figure}

In some instances, the LoTSS and DRAGONS M1 differ despite having very similar spectra. Most of these discrepancies can be explained by the polarisation bias, the offset in PI arising from noise in Stokes $Q$ and $U$. The LoTSS data have a polarisation bias of $66~\mu$Jy~PSF$^{-1}$~RMSF$^{-1}$ and $150~\mu$Jy~PSF$^{-1}$~RMSF$^{-1}$ in the northern and southern mosaics respectively \citep{erceg22,erceg24}, which in some cases overemphasises the weight of low-intensity peaks in the FD spectra. For example, the weakly polarised peak near -2~rad~m$^{-2}$ in LoTSS spectrum in Fig.~\ref{img:example_spectra_polbias}(a) pulls the M1 toward zero, in contrast to the strongly positive DRAGONS M1, which is the expected weighted mean of its broader peaks. Similarly in Fig.~\ref{img:example_spectra_polbias}(b), the multiple slightly positive LoTSS peaks contribute sufficient PI due to the polarisation bias to pull the M1 toward zero despite the strong set of peaks near -9~rad~m$^{-2}$. The example in Fig.~\ref{img:example_spectra_v2}(c) is also likely subject to this effect. \cite{VanEck2026} also pointed out challenges arising from the positive-definiteness of PI FD spectra in the context of moment calculations. We expect that correcting the polarisation bias in LoTSS would further improve the overall agreement between LoTSS and DRAGONS M1 values. 

\end{appendix}

\end{document}